\documentclass[12pt]{article}
\pdfoutput=1
\usepackage[centertags]{amsmath}
\usepackage{amsfonts}
\usepackage{amssymb}
\usepackage{amsthm}
\usepackage{graphicx}
\usepackage{etoolbox}

\newcommand{\beq}{\begin{equation}}
\newcommand{\eeq}[1]{\label{#1}\end{equation}}
\newcommand{\bea}{\begin{eqnarray}}
\newcommand{\eea}[1]{\label{#1}\end{eqnarray}}

\patchcmd{\subequations}{}%
{}{}{}

\def\a{\alpha}
\def\b{\beta}
\def\g{\gamma}

\def\d{\delta}

\def\l{\lambda}
\def\L{\Lambda}
\def\m{\mu}
\def\n{\nu}

\def\p{\pi}

\def\r{\rho}

\def\s{\sigma}

\def\vf{\varphi}

\def\o{\omega}

\def\de{\partial}
\def\nb{\nabla}

\def\bbox{\de^2}
\def\div{d\cdot\partial}
\def\grad{u\cdot\partial}
\def\spin{u\cdot d}
\def\tr{d^2}
\def\sym{u^2}

\def\ddiv{d\ast\nabla}
\def\dgrad{u\ast\nabla}
\def\dspin{u\ast d}
\def\dtr{d\ast d}
\def\dsym{u\ast u}

\begin{document}
\setlength{\topmargin}{-1cm} \setlength{\oddsidemargin}{0cm}
\setlength{\evensidemargin}{0cm}

\begin{titlepage}
\begin{center}
{\Large \bf Comments on Higher-Spin Fields\\[5pt]in Nontrivial Backgrounds}

\vspace{20pt}

{\large Rakibur Rahman$^a$ and Massimo Taronna$^b$}

\vspace{12pt}
$^a$ Max-Planck-Institut f\"ur Gravitationsphysik (Albert-Einstein-Institut)\\
     Am M\"uhlenberg 1, D-14476 Potsdam-Golm, Germany\\
\vspace{6pt}
$^b$ Universit\'e Libre de Bruxelles\\ULB-Campus Plaine C.P.~231, B-1050 Bruxelles, Belgium

\end{center}
\vspace{20pt}

\begin{abstract}
We consider the free propagation of totally symmetric massive bosonic fields in nontrivial backgrounds. The mutual compatibility of the dynamical
equations and constraints in flat space amounts to the existence of an Abelian algebra formed by the d'Alembertian, divergence and trace operators.
The latter, along with the symmetrized gradient, symmetrized metric and spin operators, actually generate a bigger non-Abelian algebra, which we refer to
as the ``consistency'' algebra. We argue that in nontrivial backgrounds, it is some deformed version of this algebra that governs the consistency
of the system. This can be motivated, for example, from the theory of charged open strings in a background gauge field, where the Virasoro algebra
ensures consistent propagation. For a gravitational background, we outline a systematic procedure of deforming the generators of the consistency
algebra in order that their commutators close. We find that equal-radii $\text{AdS}_p\times\text{S}^q$ manifolds, for arbitrary $p$ and $q$, admit
consistent propagation of massive and massless fields, with deformations that include no higher-derivative terms but are non-analytic in the curvature.
We argue that analyticity of the deformations for a generic manifold may call for the inclusion of mixed-symmetry tensor fields like in String Theory.
\end{abstract}

\end{titlepage}

\newpage
\section{Introduction}\label{sec:Intro}

It is a challenging task to construct consistent interacting theories of higher-spin (HS) fields. Generic interactions of massless fields in flat space
are in tension with HS gauge invariance, and this leads to various no-go theorems~\cite{Old,gpv,Aragone,ww,New}. Consistency issues arise even
for the free propagation of HS fields in nontrivial backgrounds as the corresponding equations of motion and constraints may cease to remain mutually
compatible, as noticed first by Fierz and Pauli~\cite{FP}. A Lagrangian formulation is free from the latter kind of difficulties, but generically suffers
from the Velo-Zwanziger problem: the resulting system of equations may allow superluminal propagation~\cite{vz,sham,kob,d3}.

A consistent Lagrangian description of free massive HS fields may be furnished by appropriate non-minimal terms, at least for backgrounds with constant
curvature~\cite{AN2,PRS,Buchbinder:2012iz,Buch,Met}. The resulting equations of motion are rather simple, but this simplicity is obscured at the
Lagrangian level~\cite{AN2}. One may wonder whether consistency can be achieved without taking any recourse to the Lagrangian formulation. Indeed, it is
possible to systematically deform the dynamical equations and constraints to render them mutually compatible in some nontrivial backgrounds~\cite{Rakib1,Rakib2}.
Moreover, consistent
interactions with other dynamical fields could be introduced more easily at the level of equations of motion. After all, nonlinear equations for interacting
HS gauge fields in AdS space have been successfully constructed~\cite{Vasiliev}, but their Lagrangian embedding is very difficult~\cite{Boulanger}.

In this article, we will restrict our attention only to totally symmetric massive bosonic fields. A spin-$s$ boson is customarily
represented by a rank-$s$ symmetric traceless Lorentz tensor, say $\vf_{\m_1\cdots\m_s}$. The dynamical equations and constraints that describe its free
propagation in flat space$-$the Fierz-Pauli equations$-$are given by:
\beq \left(\de^2-m^2\right)\vf_{\m_1\cdots\m_s}=0,\qquad\de\cdot\vf_{\m_1\cdots\m_{s-1}}=0,\qquad\vf'_{\m_1\cdots\m_{s-2}}=0,\eeq{FP}
where a dot denotes contraction of indices w.r.t.~the Minkowski metric and a prime denotes a trace. The first one of these equations is the Klein-Gordon equation
for mass $m$, while the second and third are respectively the divergence and trace constraints. The constraints are crucial in the counting of propagating degrees
of freedom. In $D$ spacetime dimensions this number is given by $\binom{D-4+s}{s}+2\binom{D-4+s}{s-1}$, which of course reduces to $2s+1$ in $D=4$.

The mutual compatibility of the Fierz-Pauli equations~(\ref{FP}) is automatic, thanks to the commuting nature of ordinary derivatives.
This is no longer true in a nontrivial background since covariant derivatives do not commute. In a constant electromagnetic background, for example,
one may consider the minimal coupling by replacing the ordinary derivatives with covariant ones, $\de_\m\rightarrow\mathcal D_\m$, to obtain
\beq \left(\mathcal D^2-m^2\right)\vf_{\m_1\cdots\m_s}=0,\qquad \mathcal D\cdot\vf_{\m_1\cdots\m_{s-1}}=0,\qquad\vf^\prime_{\m_1\cdots\m_{s-2}}=0.\eeq{FP-EM}
The mutual compatibility is lost as the equations imply an unwarranted constraint:
\beq iqF^\a{}_{(\m_1}\vf_{\m_2\cdots\m_s)\a}=0,\eeq{unwarranted}
which disappears when the background is turned off, and so the system~(\ref{FP-EM}) does not describe the same number of degrees of freedom as in flat space.

However, the covariantization~(\ref{FP-EM}) of the Fierz-Pauli equations~(\ref{FP}) is a na\"ive attempt. One may systematically incorporate
non-minimal corrections to the minimally coupled equations~(\ref{FP-EM}) in order to restore consistency~\cite{Rakib1,Kaparulin}; this systematics amounts to the
closure of an algebra generated by the deformed d'Alembertian, divergence and trace operators. In this article, we argue that a bigger algebra (including
more operators than just the d'Alembertian, divergence and trace) governs the consistency of propagation and interactions of HS fields. The trio therefore generates
a subalgebra of the latter algebra.

The organization of this article is as follows. In the remaining of this section, we explain the operator formalism that will be used throughout this article.
In Section \ref{sec:Flat}, we elucidate what we mean by the ``consistency'' algebra for free HS fields in flat space and why it is relevant. To motivate the
importance of such an algebra, we consider in Section \ref{sec:String} the theory of charged open
strings in a constant electromagnetic background, where it is the Virasoro algebra whose closure ensures consistency. We devote Section \ref{sec:Curved} to
the free propagation of HS fields in a gravitational background, where under some simplifying assumptions we outline a systematic procedure of deforming
the generators of the consistency algebra in order that their commutators close. We show among others that equal-radii $\text{AdS}_p\times\text{S}^q$ manifolds,
for arbitrary $p$ and $q$, consistently propagate totally symmetric massive bosonic fields without invoking higher-derivative kinetic terms. Moreover, these
manifolds admit the propagation of massless fields just like AdS space. We conclude in Section \ref{sec:Remarks} with some remarks and open questions.
\subsubsection*{The Operator Formalism}
In the operator formalism, contraction and symmetrization of indices are realized through auxiliary variables,
so that tensor operations are much simplified in terms of operator calculus. Symmetric fields are represented by generating
functions: \beq \vf(x,u)=\frac{1}{s!}\,\vf_{\m_1\cdots\m_s}(x)\,e^{\m_1}_{a_1}(x)u^{a_1}\,\cdots\,e^{\m_s}_{a_s}(x)u^{a_s},\eeq{field}
where $e^{\m}_a$ is the vielbein and $u^a$ is an auxiliary tangent variable. The action of the covariant derivative is defined as
a differential operator involving both $x$ and $u$: \beq \nb_\m=\bar{\nb}_\m+\o_\m{}^{ab}u_a\tfrac{\de}{\de u^b},\eeq{covD} with
$\bar{\nb}_\m$ being the standard covariant derivative acting on naked tensorial indices, and $\o_\m{}^{ab}$ the spin connection.
We will work only with the contracted auxiliary variable and the associated derivative: \beq u^\m\equiv e^{\m}_{a}(x)u^{a},\quad d_{\m}
\equiv e_{\m}^{a}(x)\tfrac{\de}{\de u^a}.\eeq{u-du} Then the vielbein postulate implies $[\nb_\m,u^\n]=0$ and $[\nb_\m,d_\n]=0$.
The commutator of covariant derivatives on a scalar function of $u$ and $d$ will be given by:
\beq [\nb_\m,\nb_\n]=R_{\m\n\r\s}(x)u^\r d^\s.\eeq{commutator}

\section{Consistency Algebra in Flat Space}\label{sec:Flat}

Let us note that the isometries of $D$-dimensional Minkowski space are captured by the Poincar\'e group $ISO(D-1,1)$, which incorporates the
momentum as generator of spacetime translations and the Lorentz generators. For flat space, where $e^\m_a=\d^\m_a$, one can construct the following set of basic
operators~\cite{Maxim}:
\beq \mathfrak{g}=\left\{\bbox,\,\div,\,\tr,\,\grad,\,\sym,\,\spin\right\},\eeq{set}
which commute with the Poincar\'e generators. The set~(\ref{set}) comprises six operators: the d'Alembertian $\bbox$, divergence $\div$, trace $\tr$, symmetrized
gradient $\grad$, symmetrized metric $\sym$, and spin $\spin$. The first three appear in the Fierz-Pauli equations~(\ref{FP}), which can now be rewritten as:
\beq \bbox\vf=m^2\vf,\qquad \div\,\vf=0,\qquad \tr\vf=0.\eeq{LWC}
The mutual compatibility of these equations can be reexpressed in terms of the following commutation relations:
\beq \left[\,\bbox,\,\div\,\right]=0,\qquad \left[\,\bbox,\,\tr\,\right]=0,\qquad\left[\,\div,\,\tr\,\right]=0,\eeq{ly3}
which imply that the d'Alembertian, divergence and trace form an Abelian algebra.

The algebra~(\ref{ly3}) is in fact a subalgebra of a non-Abelian algebra generated by all the elements of $\mathfrak g$. In particular, the d'Alembertian operator
by definition is the commutator of divergence and gradient:
\beq \left[\,\div,\,\grad\,\right]=\bbox.\eeq{comm1}
The other nontrivial commutators include:
\begin{subequations}\label{nameall}\begin{align}
\left[\,\div,\,\sym\,\right]&=2\grad,\label{comm3}\\\left[\,\tr,\,\grad\,\right]&=2\div,\label{comm4}\\
\left[\,\tr,\,\sym\,\right]&=4\spin+2D,\label{comm6}\\\left[\,\div,\,\spin\,\right]&=\div,\label{comm2}\\
\left[\,\spin,\,\grad\,\right]&=\grad,\label{comm7}\\\left[\,\tr,\,\spin\,\right]&=2\tr,\label{comm5}\\
\left[\,\spin,\,\sym\,\right]&=2\sym\label{comm8}.
\end{align}\end{subequations}

The ``consistency'' algebra we will consider is simply the set of operators $\mathfrak g$ enumerated in Eq.~(\ref{set}), given the nontrivial commutation
relations~(\ref{comm1})--(\ref{nameall}). Note that this algebra commutes with the Poincar\'e generators. In a generic manifold, one may have a different
set of isometry generators if any. Yet, it makes sense to talk about some deformed version of the operators~(\ref{set}), and require that they generate an
algebra. Moreover, the deformed d'Alembertian,  divergence and trace operators should generate a subalgebra (perhaps non-Abelian) to ensure that the deformed
Fierz-Pauli equations remain mutually compatible.

\section{Charged Open String in EM Background}\label{sec:String}

To motivate the important role played by such an algebra, we take recourse to String Theory. For a flat background, one can construct an infinite set of Virasoro
generators that commute with the target-space isometry. Of course, Poincar\'e symmetry is broken by the presence of a constant electromagnetic (EM) background, but the Virasoro
algebra prevails for charged open strings modulo deformations of the individual generators~\cite{AN2}. It is the Virasoro algebra whose closure ensures consistent propagation
of the massive HS string excitations in a constant EM background.

The world-sheet sigma model for a charged open bosonic string in a constant EM background is exactly solvable~\cite{AN2,Fradkin:1985qd}. Upon quantization, one finds the
usual infinite set of creation and annihilation operators\footnote{These operators are well defined in the regimes of physical interest.}:
\beq [\,a_m^\mu,a_n^{\dag\nu}\,]\,=\,\eta^{\mu\nu}\delta_{mn},\qquad [\,a_m^\mu,\,a_n^{\nu}\,]\,=\,[\,a_m^{\dag\mu},\,a_n^{\dag\nu}\,]\,=\,0
\qquad m,n\in\mathbb{N}_1,\eeq{crann} along with the Virasoro algebra
\beq [L_m,L_n]\,=\,(m-n)L_{m+n}+\tfrac{1}{12}\,D(m^3-m)\delta_{m,-n}\,,\eeq{r23}
which is the same as in flat space. The Virasoro generators do get deformed in the EM background. For example, one has (with $\a'=\tfrac{1}{2}$)
\begin{subequations}\begin{align}
L_0&=-\tfrac{1}{2}\mathfrak{D}^2+\sum_{m=1}^{\infty}(m+iG)_{\mu\nu}a_m^{\dag\mu}a_m^\nu+\tfrac{1}{4}\text{Tr}G^2,\label{Vir0}\\
L_1&=-i\left[\sqrt{1+iG}\right]_{\mu\nu}\mathfrak{D}^\mu a_1^\nu+\sum_{m=2}^{\infty}\left[\sqrt{(m+iG)(m-1+iG)}\right]_{\mu\nu}
a_{m-1}^{\dag\mu}a_m^\nu,\label{Vir1}\\L_2&=-i\left[\sqrt{2+iG}\right]_{\mu\nu}\mathfrak{D}^\mu a_2^\nu+\tfrac{1}{2}\left[\sqrt{1+G^2}\right]_{\mu\nu}
a_1^\mu a_1^\nu\nonumber\\&+\sum_{m=3}^{\infty}\left[\sqrt{(m+iG)(m-2+iG)}\right]_{\mu\nu}a_{m-2}^{\dag\mu}a_m^\nu,\label{Vir2}
\end{align}\end{subequations}
where $\mathfrak D^\mu$ is the covariant derivative up to a rotation,
\beq \mathfrak D^\mu=\left(\sqrt{G/qF}\right)^\m{}_\n\,\mathcal D^\nu\,,\qquad [\,\mathcal D^\m,\,\mathcal D^\n\,]=iqF^{\m\n},\eeq{covder0}
with $q=q_0+q_\pi$ being the total charge of the string, and
\beq G=\frac{1}{\pi}\left[\,\tanh^{-1}(\pi q_0F)+\tanh^{-1}(\pi q_\p F)\,\right].\eeq{covder1}

The physical state conditions for string states translate into a set of Fierz-Pauli equations for the string fields. The dynamical equations
and constraints are deformed, and their mutual compatibility is guaranteed by the Virasoro algebra~(\ref{r23}).

Restricting attention to totally symmetric fields, as we do in this article, means that we consider only the first Regge trajectory of string
excitations and exclude the subleading Regge trajectories. This is tantamount to switching off all the creation and annihilation operators but
$a_1^{\dagger\m}$ and $a_1^\m$, which leaves us with only five nontrivial Virasoro generators: $L_0$, $L_{\pm1}$ and $L_{\pm2}$. In flat space,
this quintet combines with the number operator, $\mathcal N\equiv\sum_{n=1}^\infty n a_n^\dagger\cdot a_n$, to generate what we call the
``consistency'' algebra for symmetric fields. In a constant EM background, however, this smaller set of operators no longer constitute an algebra,
and one needs to turn on all the creation and annihilation operators to construct a set of operators that do form a closed algebra$-$the Virasoro
algebra~\cite{RT}. In other words, consistency of string field theory in a constant EM background is achieved through the inclusion of mixed
symmetry fields.  Our simplified approach does not include this feature, and may therefore lead to possibilities that are not realized in
String Theory~\cite{Buchbinder:1999be}.

\newpage
\section{Consistency Algebra in Curved Manifolds}\label{sec:Curved}

In a curved background, the ordinary derivatives will be replaced by covariant derivatives: $\de_\m\rightarrow\nb_\m$, which do not commute but follow Eq.~(\ref{commutator}).
Therefore, when the operators~(\ref{set}) are na\"ively covariantized, their commutators give rise to terms proportional to the curvature tensor and its derivatives, and
so the consistency algebra ceases to close in general. In this section, we will outline a systematic procedure to deform these generators such that the consistency algebra
closes.

First, let us note that the contracted auxiliary variable $u^\m$ and the associated derivative $d_\m$ can be considered as a pair of creation and annihilation operators:
\beq [d_\m,u^\n]=\d_\m^\n,\qquad [d_\m,d_\n]=[u^\m,u^\n]=0.\eeq{cr-ann}

We start the deformation procedure with the following ansatz for the divergence:
\beq \ddiv\equiv H^{\m\n}d_\m\nabla_\n,\eeq{defdiv}
where $H^{\m\n}$ is a function of the curvature tensor and its derivatives. For simplicity, we consider neither the possible dependency of $H^{\m\n}$ on the contracted
auxiliary variable and its associated derivative nor the appearance of higher spacetime derivatives. By Hermiticity, we also have the deformed gradient:
\beq \dgrad\equiv H_\m{}^\n u^\m\nabla_\n.\eeq{defgrad}
In view of Eq.~(\ref{comm1}), we now define the deformed d'Alembertian operator $\Box$ as the commutator of the deformed divergence~(\ref{defdiv}) and gradient~(\ref{defgrad}):
\beq \Box\equiv[\,\ddiv,\,\dgrad\,].\eeq{dcomm1}
To write it more explicitly, we further make a simplifying assumption that $H^{\m\n}$ is a symmetric tensor.
Then, the deformed d'Alembertian~(\ref{dcomm1}) reads:
\beq \Box=H^2_{\m\n}\nb^\m\nb^\n+H^\m{}_\a H^\n{}_\b R^{\a\b\r\s}\left(g_{\m\r}u_\n d_\s+u_\n u_\r d_\m d_\s\right)+\cdots,\eeq{fullbox}
where the ellipses stand for terms containing derivatives of the Riemann tensor.

Unlike in flat space, the d'Alembertian operator now has a non-vanishing commutator with the divergence. It is of the form:
\beq [\,\ddiv,\,\Box\,]=X_{\m\n\r\s}\nb^\m u^\n d^\r d^\s+Y_{\m\n} d^\m\nb^\n+\cdots\,,\eeq{newdcomm1}
where the tensors $X$ and $Y$ are given by
\begin{subequations}\begin{align}
X_{\m\n\r\s}&\equiv-3H^2_{\m\a}H_{\r\b}R^{\a\b}{}_{\n\s}-H_\m{}^\a H_\n{}^\b H_\r{}^\g R_{\a\g\b\s}\,,\label{Xdefined}\\
Y_{\m\n}&\equiv\left(3H^2_{\m\r}H_{\a\b}-H_{\m\r}H^2_{\a\b}\right)R^{\a\r\b}{}_\n\,,\label{Ydefined}
\end{align}\end{subequations}
and again the ellipses stand for terms containing derivatives of the curvature.

The commutator~(\ref{newdcomm1}) must close up to the deformed divergence and a suitably deformed trace to form a deformed counterpart of the subalgebra~(\ref{ly3}).
This would ensure the mutual compatibility of the dynamical equations and constraints. Now, the form of $X$ and $Y$ suggests that the generic solution for $H$,
if any, is very non-linear and possibly non-analytic in the curvature. Since the generic problem is hard to solve, one can make a case by case study to find allowed
backgrounds that could close the consistency algebra under the given assumptions. Below we consider one particular class of backgrounds: $\text{AdS}_p\times\text{S}^q$
with equal radii but arbitrary $p$ and $q$. As we will show, such manifolds indeed close the consistency algebra.

\subsection*{$\text{AdS}_p\times\text{S}^q$ with Equal Radii}

Note that any $\text{AdS}_p\times\text{S}^q$ is a symmetric space, i.e., its Riemann tensor is covariantly constant. This immediately sets to zero all the terms denoted
by the ellipses appearing in Eqs.~(\ref{fullbox}) and~(\ref{newdcomm1}). Moreover, if the radii of $\text{AdS}_p$ and $\text{S}^q$ have the same value $l$, the manifold
is also conformally flat. The nontrivial parts of the Riemann tensor are then the traceless Ricci tensor $S^{\m\n}$ and the curvature scalar $R$. These quantities are
given by\footnote{Curiously, whenever $p=q$, the former quantity squares to unity while the latter vanishes.}
\begin{subequations}\begin{align}
S^\m{}_\n&=\frac{p+q-2}{(p+q)\,l^2}\left(-q\,\d^a_b+p\,\d^i_j\right),\label{S-given}\\
R&=-\frac{1}{l^2}(p-q)(p+q-1),\label{R-given}
\end{align}\end{subequations}
where the indices $a,b=0,1,\dots,p-1$ refer to $\text{AdS}_p$, and $i,j=1,2,\dots,q$ to $\text{S}^q$.

We now claim that the deformation tensor $H^{\m\n}$ is given by
\beq H^\m{}_\n~=~\frac{p}{p+q}\left(\d^\m_\n-\frac{(p+q)\,l^2}{p(p+q-2)}\,S^\m{}_\n\right).\eeq{Harbdef}
It is easy to see that this quantity is actually a covariant projector,
\beq H^\m{}_\r H^\r{}_\n=H^\m{}_\n,\qquad H^\m{}_\m=p,\eeq{rrrr}
which also satisfies
\beq \qquad H^\m{}_\r S^\r{}_\n=-\frac{q(p+q-2)}{(p+q)\,l^2}\,H^\m{}_\n\,.\eeq{rrrr1}
Given the properties~(\ref{rrrr}) and~(\ref{rrrr1}), one finds that the commutator~(\ref{newdcomm1}) indeed closes:
\beq \left[\,\ddiv,\,\Box\,\right]=-\frac{2}{l^2}\left((2\dspin+p-1)\ddiv-2\dgrad\,\dtr\right),\eeq{dcomm9}
where an asterisk denotes, as usual, the contraction of a pair of indices w.r.t. the covariant projector $H^{\m\n}$, and $\dtr$ is identified as the deformed trace operator.

The whole set of operators forming the consistency algebra is given by
\beq \tilde{\mathfrak g}=\left\{~\Box,~\ddiv,~\dtr,~\dgrad,~\dspin,~\dsym~\right\}.\eeq{newset}
While the commutation relation~(\ref{comm1}) is directly taken into account by the defining commutator~(\ref{dcomm1}), the relations~(\ref{comm3})--(\ref{comm8}) are deformed
respectively into
\begin{subequations}\begin{align}
\left[\,\ddiv,\,\dsym\,\right]&=2\dgrad,\label{dcomm3}\\\left[\,\dtr,\,\dgrad\,\right]&=2\ddiv,\label{dcomm4}\\
\left[\,\dtr,\,\dsym\,\right]&=4\dspin+2p,\label{dcomm6}\\\left[\,\ddiv,\,\dspin\,\right]&=\ddiv,\label{dcomm2}\\
\left[\,\dspin,\,\dgrad\,\right]&=\dgrad,\label{dcomm7}\\\left[\,\dtr,\,\dspin\,\right]&=2\dtr,\label{dcomm5}\\
\left[\,\dspin,\,\dsym\,\right]&=2\dsym\label{dcomm8}.
\end{align}\end{subequations}
The only other nontrivial commutator is the Hermitian conjugate of Eq.~(\ref{dcomm9}):
\beq \left[\,\dgrad,\,\Box\,\right]=+\frac{2}{l^2}\left(\dgrad(2\dspin+p-1)-2\dsym\,\ddiv\right).\eeq{dcomm10}

Therefore, the consistency algebra closes up to deformations of the generators that depend on the curvature. Note that the algebra makes sense even for $q=0$,
in which case $S_{\m\n}=0$, and the projector $H_{\m\n}$ reduces to the $\text{AdS}_p$ metric. The resulting algebra is simply the one for $\text{AdS}_p$
space~\cite{Hallowell,Nutma}. This is not surprising given the fact that maximally symmetric spaces do admit consistent propagation of HS fields. The new result
is that even $\text{AdS}_p\times\text{S}^q$ manifolds, with equal radii but arbitrary $p$ and $q$, do the same without invoking higher-derivative terms. In fact,
the algebra for $q\neq0$ might be considered as a covariant uplift of the $\text{AdS}_p$ algebra.

It is expected that $\text{AdS}_p\times\text{S}^q$ admits propagation of massless HS fields. To confirm this, let us consider gauge transformations of $\vf$:
\beq \d\vf=\left(u\ast\nb\right)\l,\eeq{kkkkk0}
which are on shell, i.e., the gauge parameter $\l$ satisfies
\beq \left(\Box-\m^2\right)\l=0,\qquad \left(\ddiv\right)\l=0,\qquad \left(\dtr\right)\l=0.\eeq{kkkkk}
On the other hand, the field $\vf$ itself satisfies the Fierz-Pauli equations:
\beq \left(\Box-m^2\right)\vf=0,\qquad \left(\ddiv\right)\vf=0,\qquad \left(\dtr\right)\vf=0.\eeq{tttt}
Masslessness will correspond to the appearance of gauge symmetry for some particular values of the mass parameters $\m^2$ and $m^2$. Requiring that Eqs.~(\ref{kkkkk0})
and~(\ref{kkkkk}) constitute a gauge symmetry of the system~(\ref{tttt}), one can make use of the commutation relations~(\ref{dcomm1}),~(\ref{dcomm4}) and~(\ref{dcomm10})
to arrive at the following results:
\beq \m^2=0,\qquad M_0^2\,l^2=s^2+s(p-6)-2(p-3),\eeq{mass}
where the mass parameter $M_0^2$ is related to $m^2$ through the equations:
\beq \left(\Box-m^2\right)\vf=\left(\nb\ast\nb-M_0^2\right)\vf=0.\eeq{M0-defined}
Thus a massless point $M_0^2$ does exist, as expected.

\section{Remarks \& Outlook}\label{sec:Remarks}

In this article, we have argued that the consistency of free propagation of massive HS fields in nontrivial backgrounds can be attributed to the existence of an algebra,
which we refer to as the consistency algebra.
For totally symmetric bosonic fields, this algebra is generated by six operators: the d'Alembertian, divergence, trace, symmetrized gradient, symmetrized metric,
and spin, which do get deformed in the presence of a background. Note that the consistency algebra is expected to take into account more than just the mutual compatibility
of the Fierz-Pauli equations since the latter is realized only as a subalgebra. It is possible that the incorporation of the symmetrized gradient, symmetrized metric,
and spin operators in the consistency analysis is tantamount to assuring a Lagrangian embedding of the system.

One of our results is that $\text{AdS}_p\times\text{S}^q$ manifolds, with equal radii but arbitrary $p$ and $q$, admit consistent propagation of totally symmetric
massive and massless bosonic fields without invoking higher-derivative terms. Curiously, the $\text{AdS}_5\times\text{S}^5$ solution of String Theory belongs to this
class of manifolds. However, as already emphasized towards the end of Section~\ref{sec:String}, consistency in String Theory is supposed to be realized in a different
way, perhaps with the inclusion of mixed-symmetry fields. A key feature of our result is indeed its non-analyticity in the neighborhood of flat space$-$the
deformation tensor $H^{\m\n}$, spelled out in Eq.~(\ref{Harbdef}), blows up as the quantity $S_{\m\n}S^{\m\n}$ goes to zero ($l\rightarrow\infty$). For String Theory
in $\text{AdS}_5\times\text{S}^5$ it is possible that the deformed Virasoro generators are actually smooth in the neighborhood of zero curvature, just like they
are for an EM background.

To simplify analysis, we did not consider the most general ansatz for the deformation tensor $H^{\m\n}$. Neither did we include higher-derivative kinetic terms.
In particular, $H^{\m\n}$ may depend on the contracted auxiliary variable and its associated derivative, and may contain an antisymmetric part. These possibilities
will be taken into account in some future work~\cite{RT}. It would be interesting to find what other backgrounds, if any, may not require higher-derivative
kinetic terms to admit consistent propagation of totally symmetric massive and massless fields. Closing the algebra for more generic backgrounds might
however be impossible without invoking mixed symmetry fields and/or higher-derivative kinetic terms.

One expects that the consistency algebra should also prevail beyond free theory. At the level of interactions, non-linearities will show up in the dependency of
the generators on the dynamical fields themselves. A systematic procedure for closing the consistency algebra for such non-linear deformations may shed some
light on the nature of HS interactions. We leave this as future work.

\subsection*{Acknowledgments}

We are thankful to the organizers of the ``International Workshop on Higher Spin Gauge Theories'', especially L.~Brink, and the Institute for Advanced Study of
Nanyang Technological University, Singapore for kind hospitality and support. MT is partially supported by the Fund for Scientific Research FNRS Belgium (grant FC
6369), and by the Russian Science Foundation (grant 14-42-00047) in association with Lebedev Physical Institute.

\bibliographystyle{ws-rv-van}

\begin{thebibliography}{99}

\bibitem{Old}
  S.~Weinberg,
  Phys.\ Rev.\  {\bf 135}, B1049 (1964).

\bibitem{gpv}
  M.~T.~Grisaru and H.~N.~Pendleton,
  Phys.\ Lett.\  B {\bf 67}, 323 (1977);
  M.~T.~Grisaru, H.~N.~Pendleton and P.~van Nieuwenhuizen,
  Phys.\ Rev.\  D {\bf 15}, 996 (1977).

\bibitem{Aragone}
  C.~Aragone and S.~Deser,
  Phys.\ Lett.\ B {\bf 86}, 161 (1979);
  S.~Deser and Z.~Yang,
  Class.\ Quant.\ Grav.\  {\bf 7}, 1491 (1990).

\bibitem{ww}
  S.~Weinberg and E.~Witten,
  Phys.\ Lett.\  B {\bf 96}, 59 (1980).

\bibitem{New}
  M.~Porrati,
  Phys.\ Rev.\ D {\bf 78}, 065016 (2008)
  [arXiv:0804.4672 [hep-th]].

\bibitem{FP}
  M.~Fierz and W.~Pauli,
  Proc.\ Roy.\ Soc.\ Lond.\  A {\bf 173}, 211 (1939).

\bibitem{vz}
  G.~Velo and D.~Zwanziger,
  Phys.\ Rev.\  {\bf 186}, 1337 (1969),
  Phys.\ Rev.\  {\bf 188}, 2218 (1969);
  G.~Velo,
  Nucl.\ Phys.\  B {\bf 43}, 389 (1972).

\bibitem{sham}
  A.~Shamaly and A.~Z.~Capri,
  Annals Phys.\  {\bf 74}, 503 (1972);
  M.~Hortacsu,
  Phys.\ Rev.\ D {\bf 9}, 928 (1974).

\bibitem{kob}
  M.~Kobayashi and A.~Shamaly,
  Phys.\ Rev.\  D {\bf 17}, 2179 (1978),
  Prog.\ Theor.\ Phys.\  {\bf 61}, 656 (1979).

\bibitem{d3}
  S.~Deser, V.~Pascalutsa and A.~Waldron,
  Phys.\ Rev.\  D {\bf 62}, 105031 (2000)
  [arXiv:hep-th/0003011];
  S.~Deser and A.~Waldron,
  Nucl.\ Phys.\ B {\bf 631}, 369 (2002)
  [hep-th/0112182].

\bibitem{AN2}
  P.~C.~Argyres and C.~R.~Nappi,
  Phys.\ Lett.\  B {\bf 224}, 89 (1989);
  M.~Porrati, R.~Rahman and A.~Sagnotti,
  Nucl.\ Phys.\ B {\bf 846}, 250 (2011)
  [arXiv:1011.6411 [hep-th]].

\bibitem{PRS}
  M.~Porrati, R.~Rahman,
  Phys.\ Rev.\  {\bf D80}, 025009 (2009)
  [arXiv:0906.1432 [hep-th]].

\bibitem{Buchbinder:2012iz}
  I.~L.~Buchbinder, T.~V.~Snegirev and Y.~M.~Zinoviev,
  Nucl.\ Phys.\ B {\bf 864}, 694 (2012)
  [arXiv:1204.2341 [hep-th]].

\bibitem{Buch}
  I.~L.~Buchbinder, D.~M.~Gitman and V.~D.~Pershin,
  Phys.\ Lett.\ B {\bf 492}, 161 (2000)
  [hep-th/0006144];
  I.~L.~Buchbinder, V.~A.~Krykhtin and A.~A.~Reshetnyak,
  Nucl.\ Phys.\ B {\bf 787}, 211 (2007)
  [hep-th/0703049].
  I.~L.~Buchbinder and V.~A.~Krykhtin,
  [arXiv:0710.5715 [hep-th]];
  I.~L.~Buchbinder, V.~A.~Krykhtin and P.~M.~Lavrov,
  Nucl.\ Phys.\ B {\bf 762}, 344 (2007)
  [hep-th/0608005],
  Mod.\ Phys.\ Lett.\ A {\bf 26}, 1183 (2011)
  [arXiv:1101.4860 [hep-th]].

\bibitem{Met}
  Y.~M.~Zinoviev,
  hep-th/0108192,
  Nucl.\ Phys.\ B {\bf 770}, 83 (2007)
  [hep-th/0609170],
  Nucl.\ Phys.\ B {\bf 821}, 431 (2009)
  [arXiv:0901.3462 [hep-th]].

\bibitem{Rakib1}
  I.~Cortese, R.~Rahman and M.~Sivakumar,
  Nucl.\ Phys.\ B {\bf 879}, 143 (2014)
  [arXiv:1307.7710 [hep-th]].

\bibitem{Rakib2}
  M.~Kulaxizi and R.~Rahman,
  JHEP {\bf 1410}, 193 (2014)
  [arXiv:1409.1942 [hep-th]].

\bibitem{Vasiliev}
  E.~S.~Fradkin and M.~A.~Vasiliev,
  Phys.\ Lett.\ B {\bf 189}, 89 (1987),
  Nucl.\ Phys.\ B {\bf 291}, 141 (1987);
  M.~A.~Vasiliev,
  Phys.\ Lett.\ B {\bf 243}, 378 (1990),
  Class.\ Quant.\ Grav.\  {\bf 8}, 1387 (1991),
  Phys.\ Lett.\ B {\bf 285}, 225  (1992),
  Phys.\ Lett.\  B {\bf 567}, 139 (2003)
  [arXiv:hep-th/0304049].

\bibitem{Boulanger}
  N.~Boulanger and P.~Sundell,
  J.\ Phys.\ A {\bf 44}, 495402 (2011)
  [arXiv:1102.2219 [hep-th]];
  N.~Boulanger, N.~Colombo and P.~Sundell,
  JHEP {\bf 1210}, 043 (2012)
  [arXiv:1205.3339 [hep-th]].

\bibitem{Kaparulin}
  D.~S.~Kaparulin, S.~L.~Lyakhovich and A.~A.~Sharapov,
  JHEP {\bf 1301}, 097 (2013)
  [arXiv:1210.6821 [hep-th]].

\bibitem{Maxim}
  K.~B.~Alkalaev, M.~Grigoriev and I.~Y.~Tipunin,
  Nucl.\ Phys.\ B {\bf 823}, 509 (2009)
  [arXiv:0811.3999 [hep-th]].

\bibitem{Fradkin:1985qd}
  E.~S.~Fradkin and A.~A.~Tseytlin,
  Phys.\ Lett.\ B {\bf 163}, 123 (1985).

\bibitem{Buchbinder:1999be}
  I.~L.~Buchbinder, V.~A.~Krykhtin and V.~D.~Pershin,
  Phys.\ Lett.\ B {\bf 466}, 216 (1999)
  [hep-th/9908028].

\bibitem{RT}
 R.~Rahman and M.~Taronna,
 \textit{work in progress}.

\bibitem{Hallowell}
  K.~Hallowell and A.~Waldron,
  Nucl.\ Phys.\ B {\bf 724}, 453 (2005)
  [hep-th/0505255],
  Commun.\ Math.\ Phys.\  {\bf 278}, 775 (2008)
  [hep-th/0702033],
  SIGMA {\bf 3}, 089 (2007)
  [arXiv:0707.3164 [math.DG]].

\bibitem{Nutma}
  T.~Nutma and M.~Taronna,
  JHEP {\bf 1406}, 066 (2014)
  [arXiv:1404.7452 [hep-th]].

\end{thebibliography}

\end{document}